\documentclass[tighten]{aastex63}

\usepackage{amssymb}
\usepackage{amsmath}
\usepackage []{csquotes}
\usepackage{xcolor}

\definecolor{mod}{rgb}{0,0,0}

\MakeOuterQuote{"}

\accepted{February 2022}
\shorttitle{Origin of the Uranus system: Tilting Impact}
\shortauthors{Rufu \& Canup}
\graphicspath{{./}{figures/}}

\begin{document}

\title{Co-accretion + giant impact origin of the Uranus system: Tilting Impact}

\correspondingauthor{Raluca Rufu}
\email{raluca@boulder.swri.edu}

\author[0000-0002-0810-4598]{Raluca Rufu}
\affiliation{Planetary Science Directorate, Southwest Research Institute, Boulder, Colorado, 80302, USA}

\author[0000-0002-2342-3458]{Robin M. Canup}
\affiliation{Planetary Science Directorate, Southwest Research Institute, Boulder, Colorado, 80302, USA}

\begin{abstract}
The origin of the Uranian satellite system remains uncertain. The four major satellites have nearly circular, co-planar orbits and the ratio of the satellite system and planetary mass resembles Jupiter's satellite system, suggesting the Uranian system was similarly formed within a disk produced by gas co-accretion. However, Uranus is a retrograde rotator with a high obliquity. The satellites orbit in its highly tilted equatorial plane in the same sense as the planet's retrograde rotation, a configuration that cannot be explained by co-accretion alone. In this work we investigate the first stages of the co-accretion + giant impact scenario proposed by \cite{MORBIDELLI2012737} for the origin of the Uranian system. In this model, a satellite system formed by co-accretion is destabilized by a giant impact that tilts the planet. The primordial satellites collide and disrupt, creating an outer debris disk that can re-orient to the planet's new equatorial plane and accrete into Uranus' 4 major satellites.  The needed reorientation out to distances comparable to outermost Oberon requires that the impact creates an inner disk with $\ge 1\%$ of Uranus' mass.  We here simulate giant impacts that appropriately tilt the planet and leave the system with an angular momentum comparable to that of the current system. We find that such impacts do not produce inner debris disks massive enough to realign the outer debris disk to the post-impact equatorial plane.  Although our results are inconsistent with the apparent requirements of a co-accretion + giant impact model, we suggest alternatives that merit further exploration.  
\end{abstract}

\keywords{Uranus, satellite system origin, impacts}

\section{Introduction} \label{sec:intro}
The formation of Uranus and Neptune remains poorly understood. Origin models vary significantly, from models that assume a gradual accretion of small bodies (e.g., \citealp{goldreich2004planet}), to those that invoke a late-stage giant impact phase \cite[e.g.,][]{izidoro2015accretion}, in which planets form from the merger of large protoplanetary-sized bodies. Thus, understanding their formation may provide additional constraints on the early evolution and orbital migration of the outer planets \cite[e.g.,][]{batygin2010early}. Satellite systems may provide additional constraints on the final stages of ice giant formation. 

Late giant impacts have been suggested to explain the formation of several satellites, including our Moon  \cite[e.g.,][]{canup2021origin}, Phobos and Deimos \cite[e.g.,][]{craddock2011phobos,citron2015formation,canup2018origin}, and Charon \citep{Canup:2005aa}. However, impacts seem unlikely to have produced the large satellites of Jupiter and Saturn. Instead, these likely formed by "co-accretion" within circumplanetary disks of gas and solids, created as a byproduct of the planet's accretion \citep{stevenson1986}. A circumplanetry disk supplied by an ongoing inflow of gas and solids from circumsolar orbit can accrete into satellites with a common mass fraction between the satellite system and planet of $\sim10^{-4}$ \citep{Canup2002, Canup_2006}. Neptune's large irregular satellite, Triton, was likely captured from heliocentric orbit \citep{Agnor_2006} and would have destroyed the initial satellite system, removing direct evidence of the primordial Neptunian satellites \citep{cuk2005constraints}. Dynamical analyses suggest that the primordial satellite system mass-fraction cannot have been substantially larger than $\sim10^{-4}$, and values substantially smaller than this value seem less preferred as well \citep{RufuCanup}, providing indirect evidence that Neptune's primordial satellite system had a $\sim10^{-4}$ mass ratio consistent with co-accretion as well. 

The origin of the Uranian satellite system remains poorly understood, and it is this system that may provide the most direct constraints on ice giant formation.  The four largest Uranian satellites (Ariel, Umbriel, Titania and Oberon) have near circular and co-planar orbits. Similar to the other giant planets, the total mass of the satellite system is $\sim10^{-4}$ times the planetary mass. The composition of these four satellites is $\sim$ $50\%$ rock, $50\%$ ice, consistent with solar composition material expected in a circum-Uranus co-accretion disk \citep{Canup_2006}. But unlike the other gas giants, Uranus is a retrograde rotator (obliquity of $98^\circ$), and its satellites orbit in its highly tilted equatorial plane in the same sense as its rotation. Co-accretion alone does not appear able to produce the current Uranian satellite system because gas inflow would produce a retrograde disk with respect to Uranus' rotation \citep[e.g.,][]{Lubow1999}, yielding satellites that orbit in the opposite sense to that observed. Moreover, the inner fifth largest satellite, Miranda, is likely ice-rich: its density is low, $\sim 1.17$ g/cm$^3$ (e.g., \citealp{marzari1998modelling}), and tectonic signs of endogenic activity seem inconsistent with large-scale internal porosity \citep{PappalardoSchubert2013}. The presence of an inner more ice-rich satellite is in contrast to the temperature dependence in co-accretion disk, which would tend to yield inner moons that are rock-rich compared to outer moons \citep{lunine1982formation,Canup2002}.  

Instead, it has been suggested \citep{Slattery1992} that Uranus' $98^\circ$ obliquity and its satellite system formed by a giant impact. Satellites accreted from an impact-generated debris disk would generally orbit in the same sense as Uranus' post-impact rotation \citep{Slattery1992,kegerreis2018consequences,reinhardt2020bifurcation}. However, impacts produce disks that are usually radially compact (a few Uranus radii, $R_{\rm Ur}$) compared to the semi-major axis of outermost Oberon (semimajor axis of $\sim23R_{\rm Ur}$, \citealp{jacobson2014orbits}).  Moons accreted from a compact disk could tidally evolve outward \cite[e.g.,][]{crida2012formation}, but this requires a very large tidal evolution rate compared to the estimated value for Uranus \cite[e.g.,][]{cuk2020dynamical}.  It has been recently proposed that an impact-generated disk could have viscously expanded as a vapor before the satellites accreted \citep{ida2020uranian,Woo20XX}. Impact-generated disks are also typically derived mainly from the outer layers of the impactor. Prior works suggest that producing a disk with half its mass in rock may require a pure-rock, $>3M_{\oplus}$ impactor (where $M_{\oplus}$ is Earth's mass; \citealp{reinhardt2020bifurcation,Woo20XX}), which is inconsistent with typical densities of large bodies in the outer solar system \cite[e.g.,][]{mckinnon2017origin}. Alternatively, the disk water vs. rock components would need to differentially evolve before the satellites accrete \citep{ida2020uranian,Woo20XX}.  It may be challenging to explain the current Uranian system from an impact alone. 
\begin{figure}
    \centering
    \includegraphics[width=1\linewidth]{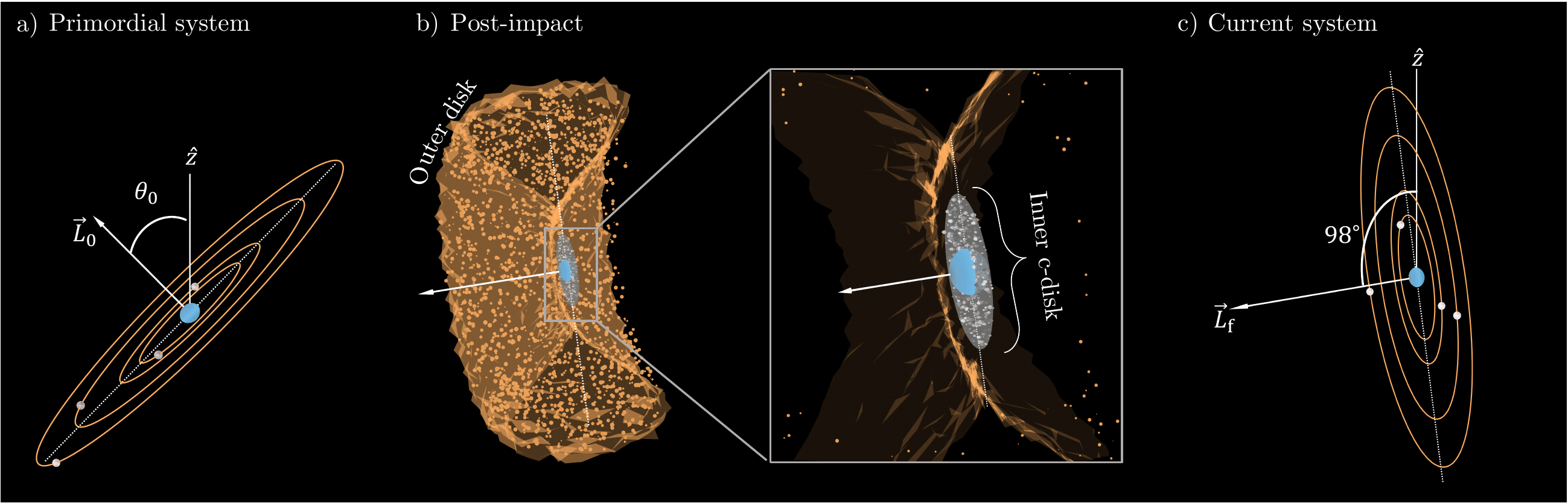}
    \caption{Schematic of the \cite{MORBIDELLI2012737} scenario. a) Initially, Uranus has a moderate obliquity ($\theta_0$) and a regular prograde satellite system formed by gas co-accretion. b) A giant impact tilts the planet to $98^\circ$ obliquity and forms an inner c-disk with mass $>10^{-2}M_{\rm Ur}$ (gray disk). The impact destabilizes the primordial satellites, causing mutually disruptive collisions and creating an outer debris cloud. The outer debris disk undergoes differential nodal regression to form a torus (orange) that is symmetric about Uranus' new equatorial plane (white dashed line). c) Uranus' satellites re-accrete from the outer disk on low inclination orbits, while essentially all of the inner c-disk and its byproducts are lost to collision with Uranus. Because the c-disk may be predominantly ice, innermost Miranda (which appears ice-rich compared to the four large outer moons) may be the largest surviving remnant of the c-disk \citep{Salmon20XXUranus}. In this work we focus on the Uranus-tipping impact to determine if this can create the needed massive inner c-disk in the \cite{MORBIDELLI2012737} scenario.}
    \label{fig:MorbiTheory}
\end{figure}

\cite{MORBIDELLI2012737} proposed a scenario that combines compelling aspects of the co-accretion and giant-impact models. In this model, Uranus initially has a moderate obliquity (similar to Neptune’s obliquity) and a prograde satellite system formed by gas co-accretion (Figure \ref{fig:MorbiTheory}-a). A giant impact tilts the planet and impulsively perturbs the primordial satellite system into mutually crossing orbits. Disruptive collisions between the satellites produce an \textit{outer debris disk} (orange disk in Figure \ref{fig:MorbiTheory}-b), while, the impact ejects material to orbits of a few $R_{\rm Ur}$, creating a compact disk, referred to as the "\textit{inner c-disk}" (gray disk in Figure \ref{fig:MorbiTheory}-b). The outer debris disk undergoes differential nodal regression to form a torus centered on Uranus' new equatorial plane. New satellites on low-inclined orbits relative to Uranus' new equatorial plane re-accrete from the outer disk. So long as the angle between the planet's pre- and post-impact spin axes is $\le 90$ degrees, the re-accreted satellites orbit in the same sense as the planet's post-impact spin.
Assuming no major losses during re-accretion, the final satellite system would then approximately preserve a $10^{-4}$ satellite system mass ratio and the primordial composition produced by co-accretion. 

\cite{MORBIDELLI2012737} showed that the current planetary oblateness ($J_2$) would randomize the ascending nodes of the outer disk out to a distance of $\sim 6R_{\rm Ur}$ (i.e., slightly beyond the orbit of Miranda). Beyond this distance, the disk's self-gravity causes the disk to precess coherently, which would result in the accretion of highly inclined outer satellites inconsistent with those observed. \cite{MORBIDELLI2012737} proposed that the Uranus tipping impact also produced an inner c-disk of mass $\sim 10^{-2} M_{\rm Ur}$ (where $M_{\rm Ur}$ is the current Uranian mass), which enhanced the effect of Uranus' $J_2$, driving node randomization in the outer disk out to Oberon's orbital distance ($\sim 23R_{\rm Ur}$). \textcolor{mod}{Further studies \citep{Salmon20XXUranus} found that the inner c-disk mass needs to be $>3 \times 10^{-3} M_{\rm Ur}$ to randomize nodes out to Oberon's distance. Because this is $> 10$ times the mass of all the current Uranian moons, nearly} 
all of the massive inner c-disk must be eventually lost, which may occur if moons spawned from the inner c-disk remained interior to the synchronous orbit ($\sim4\,R_{\rm Ur}$) and are lost to inward orbital decay.  The latter appears possible if tidal dissipation in early Uranus was intense \citep{Salmon20XXUranus}. It has been suggested that Miranda could be the largest surviving remnant of the ice-rich inner c-disk \citep{Salmon20XXUranus}.

In this work we performed impact simulations to determine whether a giant impact can both appropriately tilt Uranus and generate an inner c-disk massive enough to reorient the primordial satellite system ($>10^{-2}M_{\rm Ur}$). Previous studies \citep{Slattery1992,kegerreis2018consequences,reinhardt2020bifurcation} performed simulations of impacts onto Uranus, but they all assumed a non-rotating target. However, Uranus' pre-impact spin state is central to the viability of a co-accretion + giant impact model (see section \ref{sec:ImpactConst}), and hence we consider a pre-impact planet with a substantial initial rotation and obliquity. We constrain the allowable impact parameters by requiring that the post-impact angular momentum is comparable to the present-day Uranus system value, and that the planet's post-impact spin direction is in the same sense as the collisionally-relaxed outer debris disk.

\section{Uranus-tipping giant impact}
\subsection{Pre-impact spin and impact AM vectors}\label{sec:ImpactConst}
The requirement that satellites that re-accrete from the outer debris disk orbit in the same sense as Uranus' post-impact rotation constrains the orientation and magnitude of the impact angular momentum (AM) vector, $\vec{L}_{\rm i}$, as a function of Uranus' pre-impact AM vector, $\vec{L}_0$.

\begin{figure}
    \centering
    \includegraphics[width=0.8\linewidth]{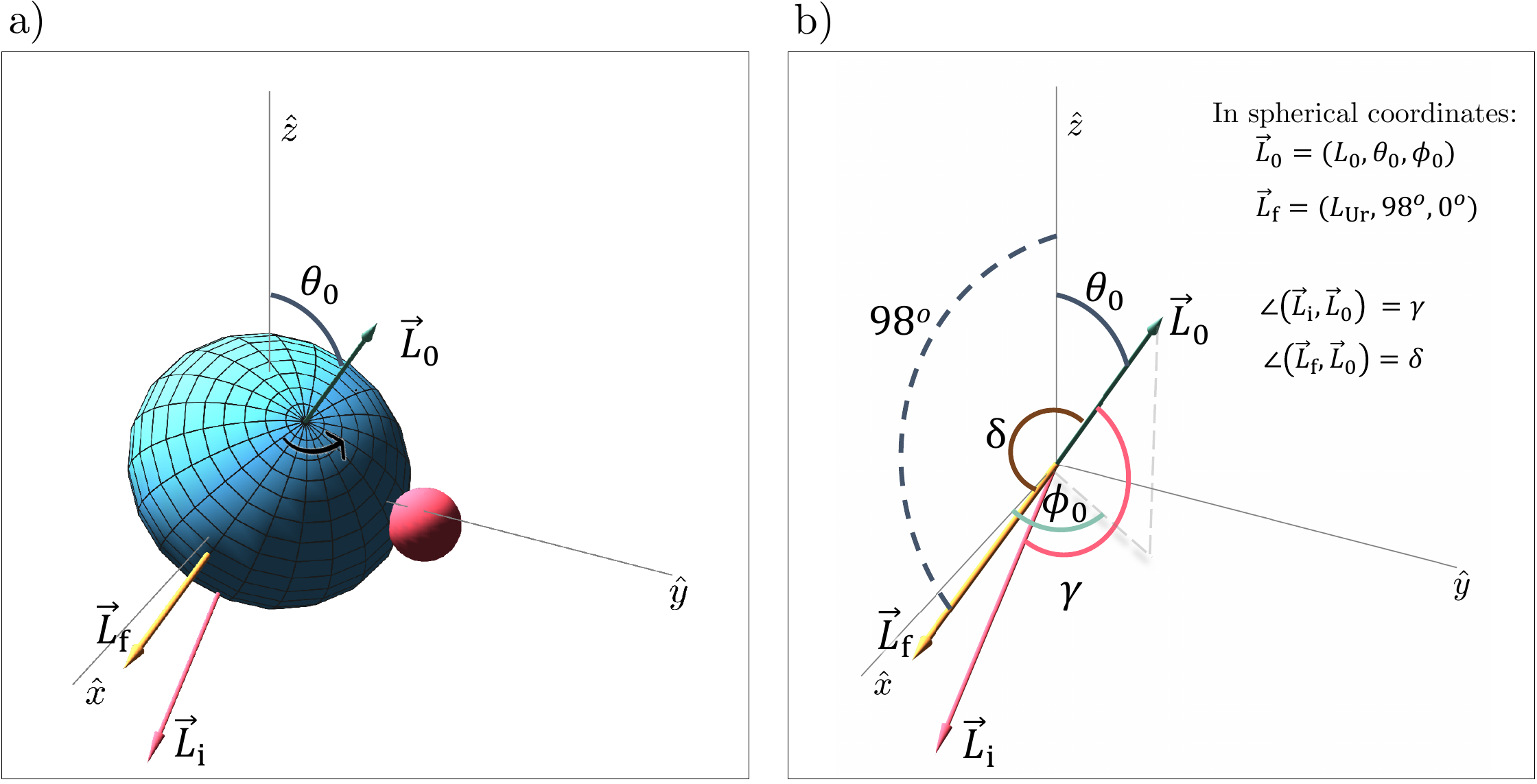}
    \caption{a) Schematic of the position and AM vectors of the proposed impact scenario. The coordinate system is defined such that the \textit{z}-axis is normal to Uranus' orbital plane, and $\vec{L}_{\rm f}$ is in the \textit{x-z} plane (i.e., in spherical coordinates the azimuthal angle is $\phi_{\rm f}=0$). The planet (blue sphere) has an initial obliquity of $\theta_0$. At the moment of impact, the impactor (red sphere) is positioned along the \textit{y}-axis.  b) Angular momentum vectors of the pre-impacting system ($\vec{L}_0$), the impact  ($\vec{L}_{\rm i}$) and post impact system ($\vec{L}_{\rm f}$).  The angle between Uranus' pre-impact AM vector and post-impact AM vector ($\delta$) must be $<90^\circ$ for final satellites to orbit in the same direction as Uranus' post-impact rotation, and must be $<60^\circ$ for outer debris to settle into orbits as distant as Oberon if the prior satellite system orbited within $10^2R_{\rm Ur}$ \citep{Salmon20XXUranus}.  The angle between $\vec{L}_0$ and $\vec{L}_{\rm i}$, $\gamma$, sets the impact plane relative to Uranus' pre- impact equator. Note that the angles $\delta$ and $\gamma$ are not necessarily in the same plane. }
    \label{fig:SchematicIS}
\end{figure}

Consider an initial Uranus with a prograde obliquity, $\theta_0$, surrounded by a prograde satellite system of mass $\sim10^{-4}M_{\rm Ur}$. The total AM of the \textit{pre-impact} system, $\vec{L}_0$, is set by Uranus' spin state (blue arrow in Figure \ref{fig:SchematicIS}-a), because the satellite system contributes minimally. The Uranus-tipping giant impact occurs by an impactor of mass $m_{\rm i}$, radius $r_{\rm i}$, velocity $V_{\rm i}$, and impact angle $\xi$ (where $\xi=90^\circ$ is a grazing collision). The impact AM vector has a magnitude \textcolor{mod}{$|L_{\rm i}|=m_{\rm i}(R_{\rm Ur}+r_{\rm i})V_{\rm i}\sin{\xi}$}, and the impact increases the planetary obliquity from $\theta_0$ to $\theta_{\rm f}=98^\circ$.  \textcolor{mod}{The impact velocity is $V_{\rm i}^2=V_{\infty}^2+V_{\rm esc}^2$, where $V_{\infty}$ is the relative velocity between the target and impactor at large separations and $V_{\rm esc}$ is the mutual escape velocity (approximately Uranus' escape velocity, $\approx 21$ km/sec). Assuming an impactor on a parabolic orbit, an upper limit for $V_{\infty}$ is $V_{\infty}=\sqrt{3}V_{\rm orb}$, where $V_{\rm orb}\sim7-8$ km/sec is the orbital velocity at $15-19$ AU \citep{nesvorny2012statistical}.} Heliocentric impactors at Uranus' orbit would thus have a low impact velocity compared to $V_{\rm esc}$, with $V_{\rm i} <1.2V_{\rm esc}$.  Thus one can approximate the collision as a perfect merger, so that the final system AM is $\vec{L}_{\rm f}\approx \vec{L}_0+\vec{L}_{\rm i}$; we will show that this is a valid approximate for most impact configurations.

For any given pre-impact planet obliquity and spin rate, one can solve for the impact AM, $\vec{L}_{\rm i}$, needed for the final AM magnitude to be comparable to that in the current Uranus system (Figure \ref{fig:SchematicIS}-b; see Appendix \ref{Appendix:Li}). There are in addition constraints on the angle ($\delta$) between Uranus' pre- and post-impact spin axes.  If the impact changed Uranus' obliquity from $\theta_0=0$ to $\theta_{\rm f}=98^\circ$, debris from collisions among the prior satellites would collisionally relax to a disk orbiting in the opposite sense as the planet's post-impact rotation, in contrast to the current satellite system. In order for outer debris to form satellites orbiting in the same sense as Uranus' post-impact rotation, the angle $\delta$ between  $\vec{L}_0$ and $\vec{L}_{\rm f}$, must be $<90^\circ$ (brown line in Figure \ref{fig:SchematicIS}-b, \citealp{MORBIDELLI2012737}). In addition, if one assumes that Uranus' original satellite system would have orbited within $10^2 R_{\rm Ur}$ (i.e., that it would be similarly radially compact as the Jovian and Saturnian regular satellites), this provides a more stringent requirement of $\delta<60^\circ$ \citep{Salmon20XXUranus} to yield a debris disk with an outer edge near Oberon's orbit as in the standard Morbidelli et al. (2012) model. The $\delta<60^\circ$ requirement implies that Uranus had a substantial pre-impact obliquity of $\ge 38^\circ$, comparable to or larger than that of present day Neptune. Uranus' initial moderate obliquity in this scenario may have been a result of a previous giant impact(s) \citep{izidoro2015accretion} and/or a spin-orbit resonance \citep{rogoszinski2020tilting}. 

\begin{figure}
    \centering
    \includegraphics[width=0.9\linewidth]{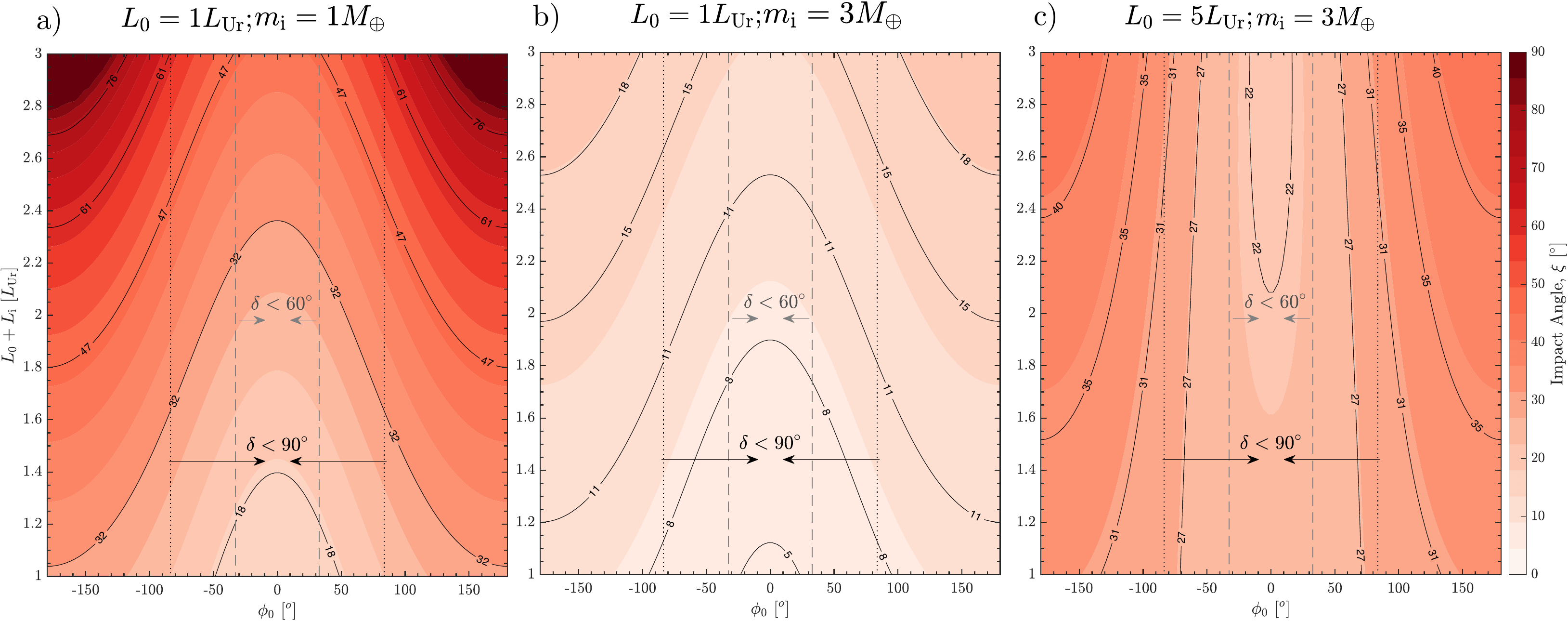}
    \caption{Impact angles that would tilt an initial Uranus from $45^\circ$ to its current obliquity (assuming that $\vec{L}_{\rm F}=\vec{L}_{0}+\vec{L}_{\rm i}$ and a, b) $L_0=1\,L_{\rm Ur}$; c) $L_0=5\,L_{\rm Ur}$) for different initial planetary AM orientation, $\phi_0$, and final AM values, $L_{\rm F}$. Here we assume an impact velocity of $1 V_{\rm esc}$ and an impactor of a) $1\, M_\oplus$, and b, c) $3\, M_\oplus$. The desired impact phase space is constrained by the vertical dotted [dashed] lines with $\delta<90^\circ$ [$\delta<60^\circ$]. Our simulations find that most of the accretionary-type impacts have $L_{\rm esc}\le 0.2L_{\rm Ur}$ (where $L_{\rm esc}$ is the AM of the escaping material), so that successful configurations that could reproduce the current Uranus system AM are located at the bottom of the plot near the $x-$axis.}
    \label{fig:AllowedImpactAngles}
\end{figure}

Figure \ref{fig:AllowedImpactAngles} shows the corresponding required impact angles for an initial obliquity of $\theta_0=45^\circ$ and $V_{\rm i} = V_{\rm esc}$ for different impactor masses ($m_{\rm i}=1,\ 3 M_\oplus$) and different initial planetary spin AM, with $L_0=1,\ 5L_{\rm Ur}$, where $L_{\rm Ur}\sim1.3\times10^{43}\ \rm{g\, cm^2\, s^{-1}}$ is the current Uranian system AM.  The implied pre-impact planet spin rate ranges from comparable to the current spin rates of Uranus and Neptune, to nearly the break-up rate for a Uranus-like planet.  The pre-impact spin orientations that can satisfy the $\delta<90^\circ$ [$\delta<60^\circ$] constraint are confined by the vertical dotted lines [vertical dashed lines]. This plot assumes perfect merger; if material and angular momentum escapes during the impact ($\vec{L}_{\rm f}<\vec{L}_0+\vec{L}_{\rm i}$), then successful cases are found by moving upward along the $y$-axis in Figure \ref{fig:AllowedImpactAngles}, allowing for an increased impact angle and a more grazing impact. 

\textcolor{mod}{For a fixed impactor mass, the maximum mass placed into orbit during a low velocity impact occurs when the scaled impact parameter is between $\sim 0.5$ to $0.7$.  For this optimal range of impact parameter, the orbiting mass generally increases as the impactor mass is increased.  Impactors with mass $\le 1M_{\oplus}$ within this impact parameter range can produce a final system with the appropriate AM, but we find that their disks are consistently too low in mass to meet the requirement described in the prior paragraph.  Substantially larger impactors would produce more massive disks.  However,} for large impactors ($3M_\oplus$) and initial planet rotational AM comparable to the current Uranus ($1L_{\rm Ur}$), the impact angles are restricted to almost head-on configurations ($\xi<15^\circ$) to yield an appropriate final AM system, which we find typically result in very low mass circumplanetary disks. For these cases, larger impact angles would tilt the planet even further than the required $\theta_{\rm f}=98^\circ$ and increase the planetary AM well beyond the current value.  In order to explore the effect of larger impact angles with $m_{\rm i}=3M_{\oplus}$ impactors, we also consider the case of rapid initial planet rotation in the opposite sense to the spin imparted by the impact itself to reduce the large impact AM (i.e., $\vec{L}_0$ that is substantial and retrograde compared to $\vec{L}_{\rm i}$). \textcolor{mod}{We note that for lower impactor masses with very grazing impacts the disk mass is expected to be small, because of the low impact energy and also because at high angles the impactor will graze and escape the planet (hit-and-run) \cite[e.g.,][]{Rufu:2017aa}.}

\subsection{Methods}
We use Smoothed Particle Hydrodynamics (SPH) to simulate impacts into Uranus, using the astrophysical code Gadget2 \citep{Springel2005} \textcolor{mod}{ that includes code modifications to incorporate a tabulated equation of state \citep{Marcus2009,marcus2011role} (available in the supplemental material of \citealp{cuk2012making})}. SPH mimics material as spherically symmetric particles, with the spatial distribution of each particle defined by a spline density weighting function, known as the kernel, and a characteristic radius, known as the smoothing length. The kinematic state of each particle evolves due to gravity, compressional heating/expansional cooling, and shock dissipation. The equation of state accounts for phase changes and includes different phases (liquid and vapor) within an SPH particle, assuming phase equilibrium.

We simulate impacts between a Uranian-like planet and four impactors ($m_{\rm i} = 0.5,\ 0.8,\ 1,\ 3 M_\oplus$). We assume the impactors are differentiated and have a $40\%$ rock and $60\%$ ice composition \citep{Lunine1993origins}. By comparison the largest ice-rock satellites of Jupiter and Saturn, have roughly 50\% rock, while outer Triton and other large KBO's may have $\sim70\%$ rock \citep{bierson2019using}. The late accretion stage of the ice giants may involve multiple giant impacts, and as discussed in previous section, a pre-impact obliquity is a crucial requirement for the needed realignment of the primordial satellites \citep{MORBIDELLI2012737}. We assume that the target has a pre-impact rotational AM of 0.5,\ 1.0,\ 1.7,\ 2.9,\ or $5L_{\rm Ur}$.  \textcolor{mod}{The latter is a limiting case of approximately the fastest rotation we could impart to the target (see below) before it would be rotationally unstable.}

We assume that the mass of the target is equal to the current Uranian mass, $M_{\rm Ur}$ and that it is differentiated containing either two or three distinct layers. \textcolor{mod}{We gradually introduce the rotation to the target and simulate it in isolation for 10 hr to allow for initial relaxation and to establish gravitational equilibrium}. For the moderate rotating planets ($L_0=0.5,\ 1,\ 1.7,\ 2.9\ L_{\rm Ur}$) we assume a rocky core ($2.3$ $M_\oplus$), a water mantle ($11.6M_\oplus$), and a hydrogen envelope ($0.86M_\oplus$), similar to previous impact studies \cite[e.g.,][]{kegerreis2018consequences}. There is considerable uncertainty in the composition and structure of Uranus (\citealp{nettelmann2013new}; see also \citealp{vazan2020explaining} who suggested that Uranus might not be completely  differentiated). The core size used in these simulations is larger than the value estimated by \cite{nettelmann2013new}, but overall this would not alter the overall impact dynamics considered here. The $1L_{\rm Ur}$ target has a normalized moment of inertia comparable to the estimated Uranian value \citep{nettelmann2013new}. \textcolor{mod}{For the limiting case of an extremely fast rotating planet with $5 L_{\rm Ur}$, the centrifugal force in the outer edges of the envelope is larger than the gravitational force (the gas envelope is not stable) for the three layer structure.  In this case, we instead adopt a two layer structure: a rocky core ($2.1 M_\oplus$) and a water (vapor) mantle ($12.2 M_\oplus$). Because this structure has a larger gyration constant (moment of inertia factor) than the 3 layer model, it can achieve a somewhat larger AM before becoming rotationally unstable. The $L_{\rm 0} = 5 L_{\rm Ur}$ state represents approximately the highest possible AM in the pre-impact planet, which when oriented in the opposite direction as that of the impact itself, i.e., with $(\vec{L}_{\rm i} \cdot \vec{L}_{\rm 0})/(L_{\rm i}L_{\rm 0}) = -1$, allows for the maximum impact parameter for the largest, $3M_{\oplus}$ mass impactor given the requirement that $\vec{L}_{\rm i} + \vec{L}_{\rm 0} \approx \vec{L}_{\rm Ur}$.}

We employ a tabulated equation of state (EOS) and characterize the rocky material by the forsterite semi-analytical EOS M-ANEOS \citep{melosh2007hydrocode}, the icy planetary mantle by the $\rm{H_{2}O}$ five-phase EOS \citep{senft2008impact}, and the outer atmosphere by the ideal gas EOS with the mean molecular weight equal to the mass of the hydrogen atom \citep[similar to][]{reinhardt2020bifurcation}. The simplified treatment of the envelope underestimates the gas densities close to the mantle-atmosphere boundary.

We performed simulations using $\sim 5\times10^5$ particles, which would sufficiently resolve a debris disks of $0.01M_{\rm Ur}$ with $\sim 5000$ particles. An additional increase in the number of SPH particles is not expected to significantly change the resulting disk masses \citep{kegerreis2019planetary}. 
The initial positions of the bodies are calculated by integrating the positions and velocities of the bodies at contact backward in time to a distance of $1.5\times$ the radii sum using a 2-body Runge-Kutta $4^{\rm th}$ order integration. 

Henceforth we will refer to the "c-disk" from the \cite{MORBIDELLI2012737} model as simply the "disk", and note that the outer debris disk is not included in our SPH simulations (but see discussion).

\subsection{Impact Analysis}
After each simulation we follow an interactive procedure  \citep{canup2001scaling} to classify the particles according to their AM and energy. Given an initial guess of the planetary mass,  $M_{\rm planet}$, the bounded particles are found.  We further classify the bound particles according to their semimajor axis equivalent, $a_{\rm eq}=l_z/\sqrt{G M_{\rm planet}}$ (where $l_z$ is the specific AM magnitude normal to the post-impact equatorial plane of the planet, which is approximately conserved in the subsequent dynamical mutual interactions between ejected particles). Disk [planet] particles are defined as those with $a_{\rm eq}>R_{\rm planet}$ [$a_{\rm eq}<R_{\rm planet}$], where the radius of the planet is calculated assuming the current Uranian density, $1.27\ \rm{g/cm^3}$. After each iteration the coordinate system is rotated such that the AM vector of the planet, $\vec{L}^\prime_{\rm planet}$ \textcolor{mod}{(defined as the sum of the angular momenta of the particles defined as being within the planet)}, is aligned along the $\hat{z}$ direction. The particle classification step is repeated with the new estimate for $M_{\rm planet}$ until convergence is achieved (usually a few iterations). The final planetary tilt is calculated using the original coordinate system such that $\theta_{\rm f}=\arccos\left(\vec{L}_{\rm planet}/\left|\vec{L}_{\rm planet}\right|\cdot\hat{z}\right)$. 
The above method for calculating the mass of the orbiting disk ignores the role of pressure support for vaporized material (see Appendix \ref{Appendix:VaporDisk}). 

We note that \cite{kegerreis2018consequences} defines orbiting mass as that which is bounded to the planet and instantaneously located beyond a distance of $1.5R_{\rm Ur}$.  Because this includes eccentric material that lacks sufficient AM to stably orbit above the planet's surface, their estimates will tend to overestimate disk mass compared with ours.  

We performed $\sim80$ simulations of different initial conditions.  Initially we simulated the impacts for 24 hours. If the disk mass after 24 hours was $>3\times10^{-5}M_{\rm Ur}$ or if large clumps expected to impact the planet are present ($\sim0.1M_{\oplus}$), we continued the simulations for an additional 48 hr or until clump impact with the planet has occurred. 


\section{Results}
\begin{figure}
    \centering
    \includegraphics[width=1\linewidth]{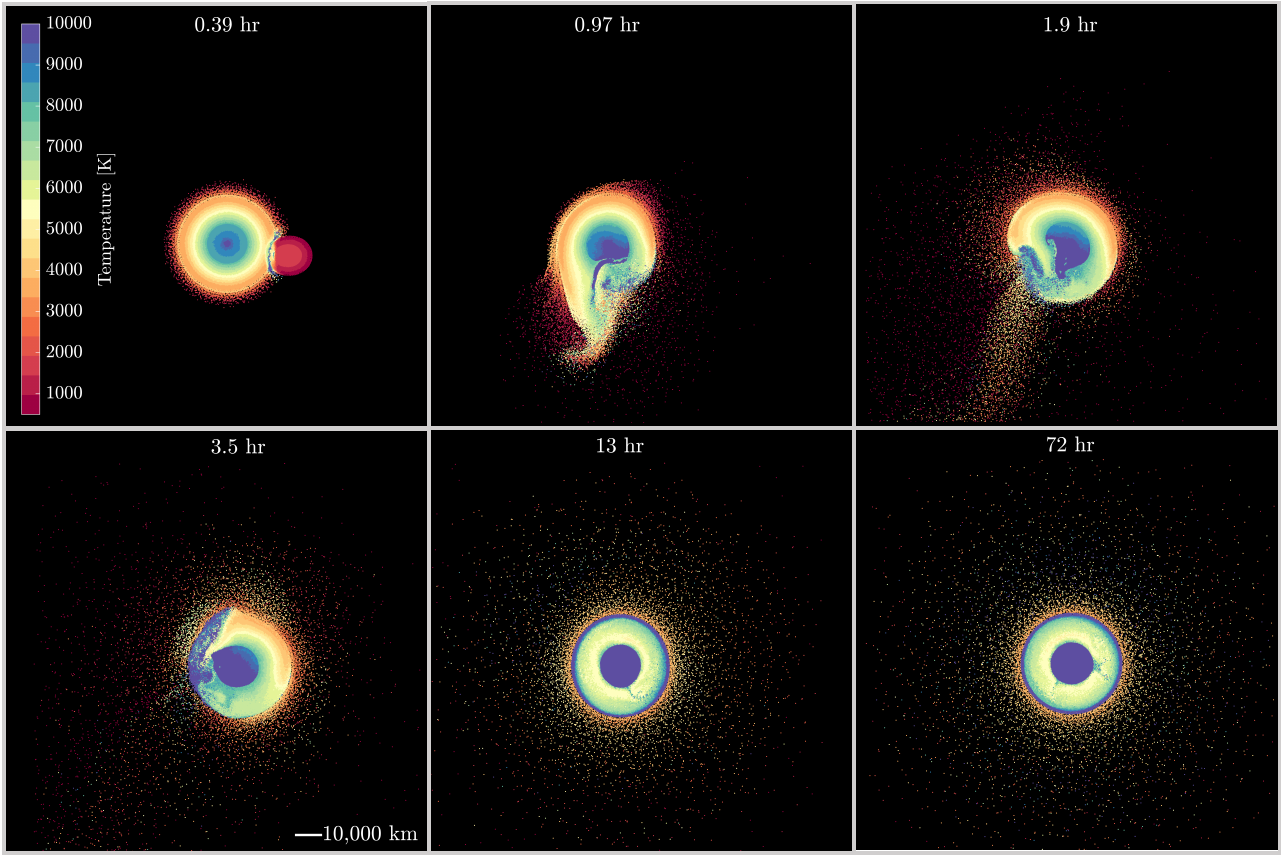}
    \caption{
    Several snapshots of a $0.8M_{\oplus}$ body impacting a Uranus-mass target with an initial rotational AM of $0.5 L_{\rm Ur}$. The colors represent the temperature of the material. This is an example of an accretionary impact, with a velocity of $V_{\rm i}= 1\ V_{\rm esc}$ and a relatively head-on collision with a $24^\circ$ impact angle. All projections are on the equatorial plane of the pre-impact target with one hemisphere removed. The resulting disk mass is $3.2\times 10^{-4} M_{\rm Ur}$, and its constituents are water and H-He ($46\%$ and $54\%$ respectively). While the post-impact planetary AM, mass and planetary obliquity are similar to the current Uranian values ($1.03\ L_{\rm Ur}$, $1.06\ M_{\rm Ur}$, $94\deg$), the resulting disk mass is much less than that needed to realign the debris of the primordial satellite system to the post-impact planet's new equatorial plane \citep{MORBIDELLI2012737,Salmon20XXUranus}. Broadly similar outcomes are seen in all of our simulations that produce final systems consistent with the current Uranian system angular momentum.}
    \label{fig:Snapshot_Impact}
\end{figure}

Figure \ref{fig:Snapshot_Impact} shows a time series of \textcolor{mod} {one of the simulations of }a $m_{\rm i}=0.8 M_{\oplus}$ body impacting a Uranus-like planet with a pre-collision rotational AM $L_0 = 0.5L_{\rm Ur}$. The oblique impact (impact angle $\xi=24^\circ$) tilts the planet from an initial obliquity of $\theta_0=35^\circ$ to a final obliquity of $\theta_{ \rm f}=93.5^\circ$, and the final AM of the resulting bound system is $1.04 L_{\rm Ur}$, broadly consistent with the current Uranian system. The post-impact planet has envelope temperatures $>5000$ K. The resulting disk mass (i.e., gravitationally bound material with sufficient AM to have a circular orbit above the planet's ``surface'', where the latter is defined by the current Uranus density) is = $3.2 \times 10^{-4} M_{\rm Ur}$, vastly smaller than the $\ge 10^{-2}M_{\rm Ur}$ c-disk mass required value to realign the outer debris disk out to distances consistent with low-inclination Oberon. \textcolor{mod}{ This example simulation is representative of the majority of simulations conducted this study, which will be discussed in following sections.}

We note that, differently from the well-studied canonical Moon-forming impacts \cite[e.g.,][]{canup2004simulations}, the resulting disk and planet AM are not necessarily aligned after $72$ hours post-impact. This occurs because the impact and initial planetary AM are misaligned (see Figure \ref{fig:SchematicIS}), and the disk contains a disproportionate amount of impactor material compared to the planet. Subsequent nodal precession and inelastic collisions among the orbiting material would yield an equatorial disk extending out to a few to perhaps 10$R_{\rm Ur}$ (with the latter approximately the maximum $a_{\rm eq}$ value for the bound orbiting material), while material having $a_{\rm eq}<R_{\rm Ur}$ may fall into the planet on an orbital timescale.

\subsection{Impact-generated disk mass} \label{sec:DiskMass}

\begin{figure}
    \centering
   \includegraphics[width=1\linewidth]{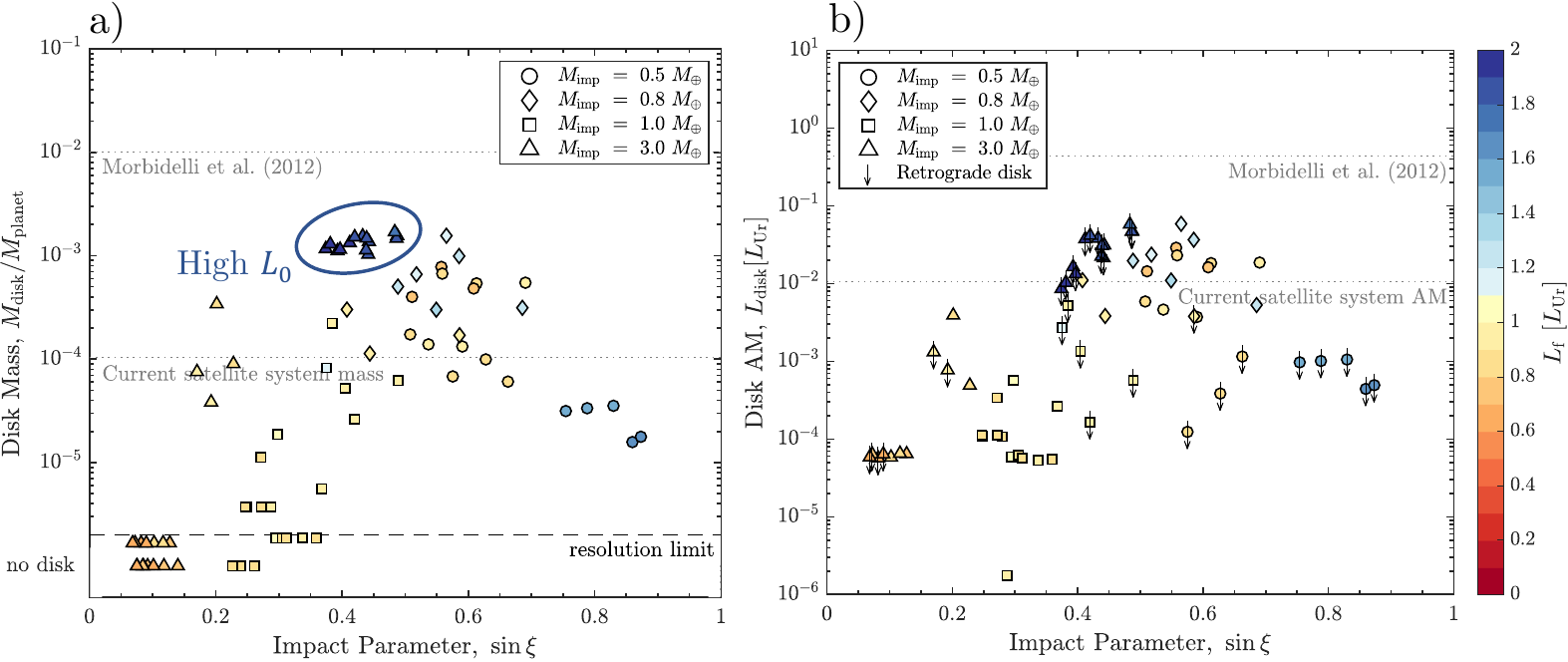}

    \caption{  a) Disk mass and b) disk AM as a function of the scaled impact parameter, $\sin{\xi}$, for impactor masses of $0.5$ (circle), $0.8$ (diamond), $1$ (square) and $3\, M_\oplus$ (triangles). \textcolor{mod}{These simulations all had impact velocities within about 20$\%$ of Uranus' escape velocity.}  The colors of the markers correspond to the post-impact bound system AM, $L_{\rm f}$, which has been normalized by a factor of $\left (M_{\rm Ur}/M_{\rm planet}\right )^{5/3}$ to compensate for the small differences in the resulting planetary mass, $M_{\rm planet}$, compared to $M_{\rm Ur}$.  Thus, successful cases require $L_{\rm f} \approx L_{\rm Ur}$ (yellow).  Points that lie below the black dashed line in panel a) are cases that did not produce disks. Dark blue points correspond to simulations with large impactors that assumed \textcolor{mod}{approximately the maximum possible planetary rotation before the impact, $L_0=5L_{\rm Ur}$.  Even when this pre-impact spin is in the opposite sense of the impact so as to maintain $\vec{L}_{\rm i} + \vec{L}_{\rm 0} \approx \vec{L}_{\rm Ur}$, these consistently leave systems with far too much final AM to be consistent with current Uranus ($L_{\rm f} \gg L_{\rm Ur}$), due to substantial escape of impactor material.} The lower grey dotted line represents the a) mass and b) AM of the current Uranian satellite. The upper grey dotted line represents the minimum disk mass (a) and AM (b) required to realign the outer disk to the planet's post-impact equatorial plane, based on analyses performed by treating the inner disk as satellite on a circular orbit at $3R_{\rm Ur}$ \citep{MORBIDELLI2012737,Salmon20XXUranus}. Arrows in panel b) indicate cases in which the c-disk is retrograde compared to the post-impact planetary rotation.}
    \label{fig:DiskMass_Lf}
\end{figure}

Figure \ref{fig:DiskMass_Lf} shows resulting disk masses and system AM values from our suite of SPH simulation. The initial parameters of the impact (see Appendix \ref{Appendix:Li}) were calculated by setting $1.1 \le (L_0+L_{\rm i})/L_{\rm Ur} \le 1.6$, which assumes small-to-modest loss of angular momentum in escaping material, and by constraining the maximum planetary tilt allowed, $\delta<90^o$.  Most of the resulting post-impact systems have final AM values that are comparable with the current system (yellow colors), indicating that most of the impacts are nearly perfect mergers \textcolor{mod}{(including some graze-and-merge impacts)}. 
Partial mergers do occur in some cases when the impactor grazes the target and a significant part of its mass escapes the system, either intact (hit-and run impact - light blue circles in Figure \ref{fig:DiskMass_Lf}) or not (disruptive hit-and run - dark blue triangles in Figure \ref{fig:DiskMass_Lf}). As discussed in section \ref{sec:ImpactConst} (also in Figure \ref{fig:AllowedImpactAngles}), higher initial rotations \textcolor{mod}{that are in the opposite sense of the impact-induced rotation} allow for more grazing impactors if perfect merger is assumed, but the SPH simulations show that the perfect merger assumption in this case is not valid, as a substantial part of the impactor and its angular momentum escapes the system.  Thus in these cases the planetary rotation is not reduced to a value that is comparable with the current system, and the final systems have large AM excesses.

Moreover, a significant fraction of the resulting disks have a total AM that is retrograde compared to the post-impact planetary rotational axis (downward arrows in Figure \ref{fig:DiskMass_Lf}-b). This occurs for all of the very fast initial rotating planets, due to the large misalignment between the impact AM and initial planetary AM. An inner retrograde moon could potentially accrete from such a disk and would tidally evolve inward and be lost.  A retrograde inner c-disk could not be the source of material that ultimately accreted into Miranda and the other inner smaller Uranian moons (e.g., \citealp{hesselbrock2019three}), as these all orbit in the same sense as Uranus' rotation.

\textcolor{mod}{Although our simulations span a wide range of impactor masses and impact conditions, including those designed to attempt to maximize the orbiting mass, none of the resulting disks are massive enough to produce nodal randomization and realignment of the outer disk to the new equatorial plane out to distances comparable to Oberon's current orbit \citep{MORBIDELLI2012737,Salmon20XXUranus}.  In our suite of simulations, impacts between an extremely fast rotating planet ($L_0=5 L_{\rm Ur}$) and a $3M_{\oplus}$ impactor resulted in the most massive disks ($\sim 2\times10^{-3}M_{\rm Ur}$).} \textcolor{mod}{The next most massive disks were produced by smaller $0.8M_{\oplus}$ impactors with a scaled impact parameter near $\sim 0.6$ to 0.5, and slower pre-impact rotation in the planet ($L_{\rm 0} = 0.5$ or 1$L_{\rm Ur}$).}  None of these \textit{c}-disks are massive enough to realign the outer disk to Oberon's distance.  In the case of the $3M_{\oplus}$ impactor, the post-impact AM values are in addition $\sim2L_{\rm Ur}$, far too high to be consistent with the current Uranian system. Overall, impacts that produce a final Uranus with an appropriate angular momentum produce \textit{c}-disks that are typically one or two orders-of-magnitude too low in mass for the \cite{MORBIDELLI2012737} model. 

\textcolor{mod}{We note that a more massive disk can be produced by a large, oblique impactor into a non-rotating planet.  For example, disk masses between $\sim 0.03$ to $0.05M_{\rm Ur}$ were produced in \cite{reinhardt2020bifurcation} by $3M_{\oplus}$ impactors, but those cases left inappropriately high-AM final systems with very large AM excesses ($>3L_{\rm Ur}$), as shown in Figure \ref{fig:CompAM}.  Similarly massive disks are also considered in \cite{Woo20XX}, but those also were produced by impacts that would have left a system with a great AM excess. Our analyses here, while focused on the \cite{MORBIDELLI2012737} model, highlight that any model involving a Uranus tipping giant impact needs to be consistent with the observed Uranus system angular momentum.}

\subsection{Disk composition and structure}

\begin{figure}
    \centering
    \includegraphics[width=0.50\linewidth]{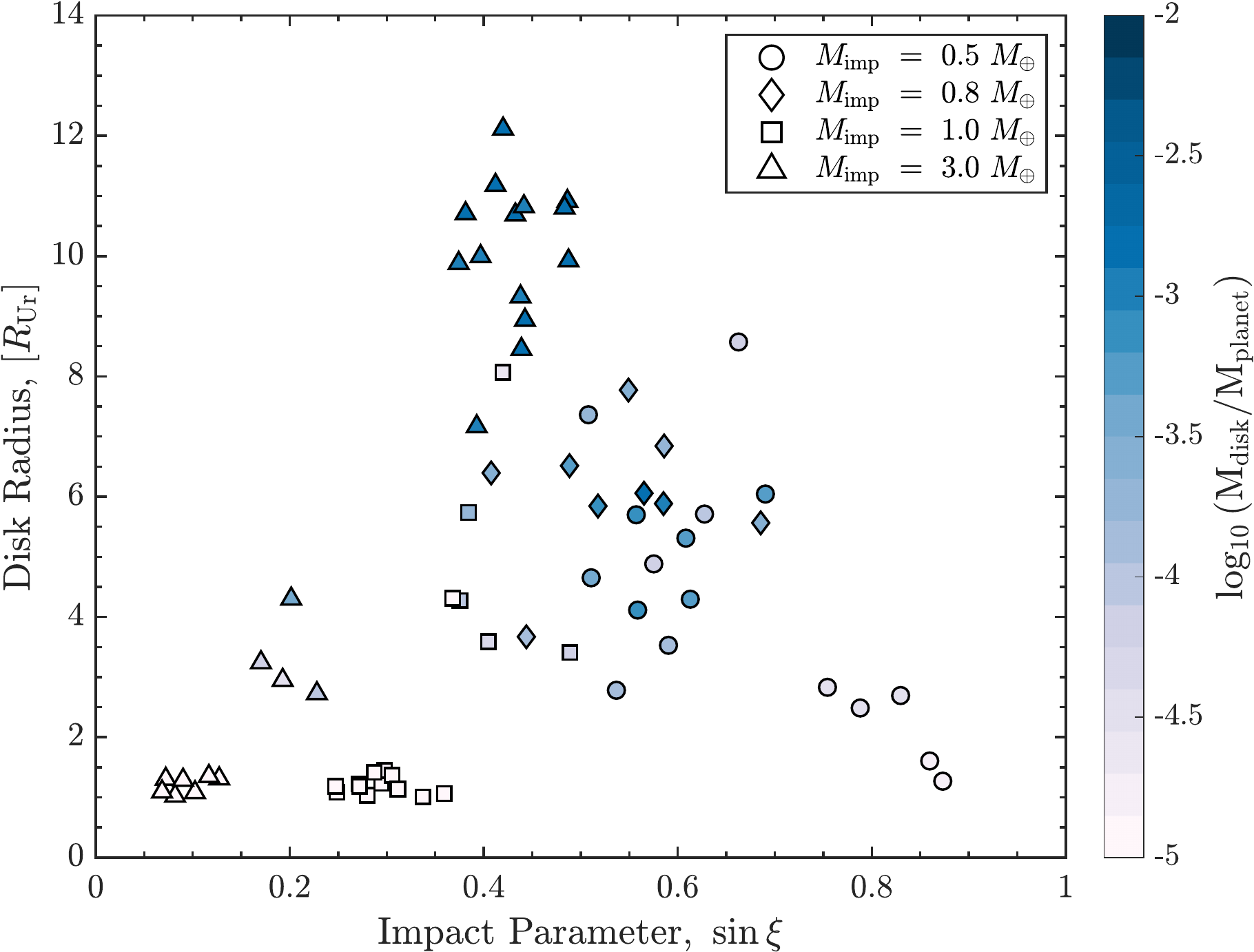}
    \caption{Disk radius as a function of impact parameter for $0.5$ (circle), $0.8$ (diamond), $1$ (square) and $3\, M_\oplus$ (triangles) impactor masses. Shown is the maximum equivalent circular orbit of bound orbiting material, neglecting pressure support (see text).  The colors of the markers correspond to the mass of the disk normalized by the current mass of Uranus. The most massive disks extend to orbits that are substantially beyond Ariel's current orbit ($7.5R_{\rm Ur}$), while the less massive disks are typically more compact in radial extent.}
    \label{fig:DiskRadius}
\end{figure}

\begin{figure}
    \centering
    \includegraphics[width=0.50\linewidth]{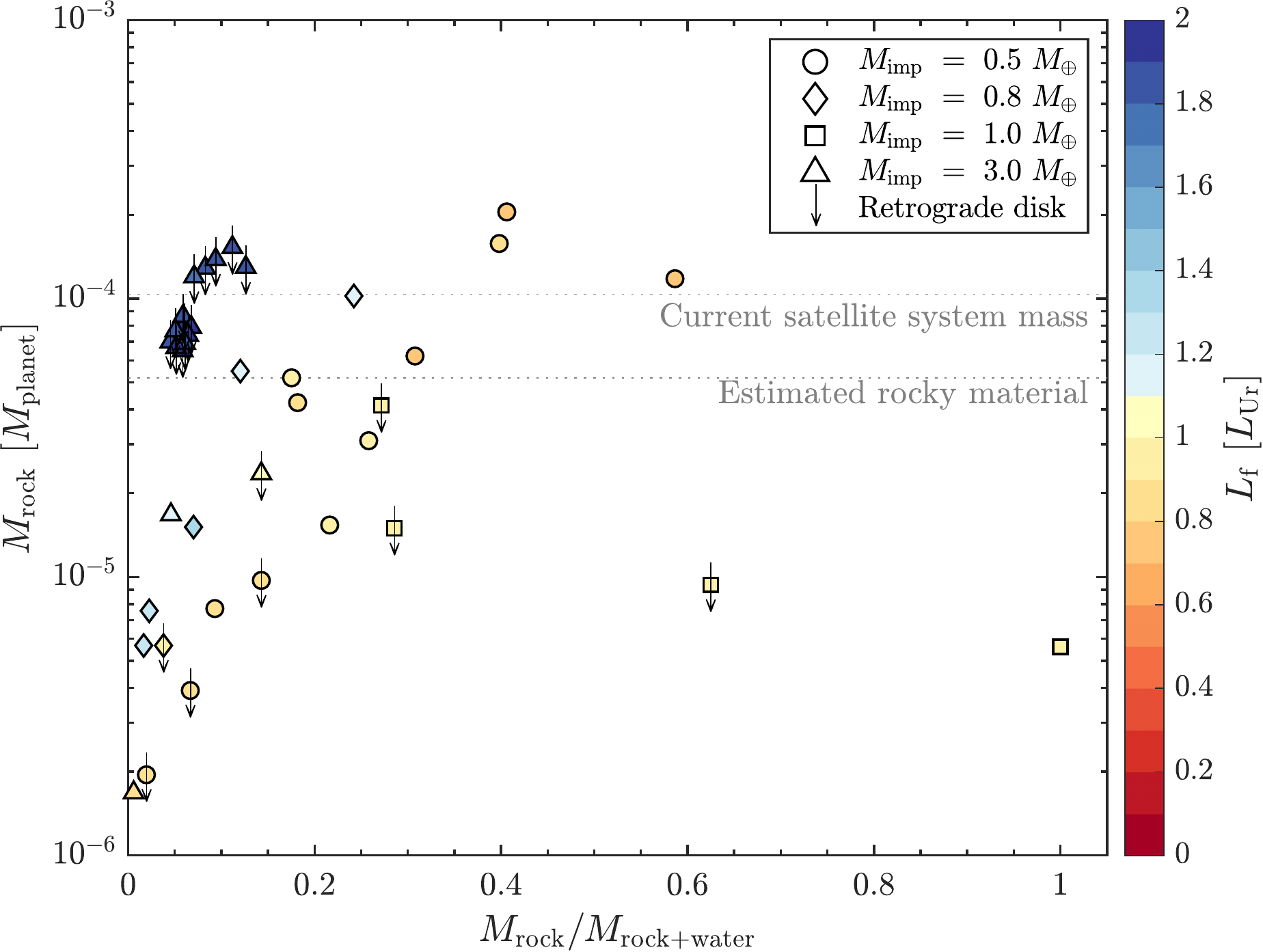}
    \caption{Rock mass in the disk as a function of the disk rock-to-total disk solids (rock + water) mass fraction for $0.5$ (circle), $0.8$ (diamond), $1$ (square) and $3\, M_\oplus$ (triangles) impactor masses. The colors of the markers correspond to the final AM of the system, $L_{\rm f}$ which has been normalized by a factor of $M_{\rm Ur}/M_{\rm planet}^{5/3}$. The upper [lower] grey dotted line shows the mass of the current Uranian satellites [the total estimated rocky mass in the Uranian satellites]. Impact-generated disks that do not contain any rocky material are not shown in this figure. Downward arrows represent cases where the disk AM is retrograde compared to the post-impact planetary rotation.} 
    \label{fig:DiskComposition}
\end{figure}

The most massive disks in our suite of simulations extend beyond $7.5R_{\rm Ur}$, Ariel's semi-major axis (Figure \ref{fig:DiskRadius}), according to their maximum equivalent semi-major axes (neglecting pressure forces). The lower, $10^{-4}M_{\rm Ur}$ mass disks are somewhat more compact, but are also not generally confined to within the Roche limit as considered in \cite{MORBIDELLI2012737} and \cite{Salmon20XXUranus}.  Thus inner disk material would be expected to mix with portions of the outer debris disk.  In most cases, the impact-generated disks are mainly derived from the outer layer of the planet and impactor, and hence they are typically rock-poor (Figure \ref{fig:DiskComposition}; note that 16 simulations did not contain any rocky material in the disk and are not shown in this figure). Contamination from inner disk material would thus increase the ice-to-rock ratio of the outer disk.  This contamination could substantially influence the outer satellite compositions, because even c-disks that are well below the mass needed to re-align the outer debris ("low mass disks") are still often comparably or more massive than the current outer satellites.  

Portions of the outer regions of the vaporized impact-generated disks are predicted to be Rayleigh unstable (see Appendix \ref{Appendix:VaporDisk}), which would lead to rapid radial redistribution of disk material \citep{Nakajima2014259}. Simulations of the evolution of vapor-rich disks (that include e.g., condensation, turbulence and accretion) are required in order to more accurately estimate the disk mass and radial distribution after it reaches a stable configuration, but the overall effect of such evolution would likely be a net mass transfer inward onto the planet, decreasing the disk mass compared to the values in Figure \ref{fig:DiskMass_Lf}-a. 

While we focus on the co-accretion + giant impact model, an alternative concept is that the current Uranian satellite system formed entirely from the disk produced by a Uranus-tipping impact, which viscously spread to orbits comparable to the Uranian satellites before it condensed and satellite accretion occurred \citep{ida2020uranian,Woo20XX}. During this expansion, an initial disk much more massive than the current Uranian moons ($10^{-2} M_{\rm Ur}$, with this value depending on the viscosity assumed) is lost while only a small part of it accretes to form the current satellites. The simulations here show that although a Uranus-tipping impact can produce a disk whose total rock mass is comparable to that in the Uranian satellites, the disks overall are typically rock-poor, and thus substantial water relative to rock would need to be lost to yield the current $\sim50\%$ rock, $50\%$ ice satellites.

\section{Discussion}

The co-accretion+giant impact scenario proposed by \cite{MORBIDELLI2012737} combines the advantages of the co-accretion model (e.g., obtaining a satellite system with the $10^{-4}$ mass ratio, 50/50 rock/ice composition) and the giant impact model (e.g., satellites orbiting in the same sense as the planet's rotation and the high planet obliquity). In this hybrid model, a satellite system formed by co-accretion is destabilized by a giant impact that tips the planet's rotation.  The primordial satellites collide and disrupt, creating an outer debris disk that is initially inclined to the Uranus' new, post-impact equatorial plane. In order for this outer debris disk to be appropriately re-aligned with Uranus' new equatorial plane out to distances consistent with Oberon, the giant impact must produce an inner massive disk containing $\ge 0.01M_{\rm Ur}$ \citep{MORBIDELLI2012737,Salmon20XXUranus}.  Here we explore whether a giant impact could both appropriately tilt the planet and produce an impact-generated inner ``c-disk'' massive enough to realign the outer disk.

\begin{figure}
    \centering
    \includegraphics[width=0.45\linewidth]{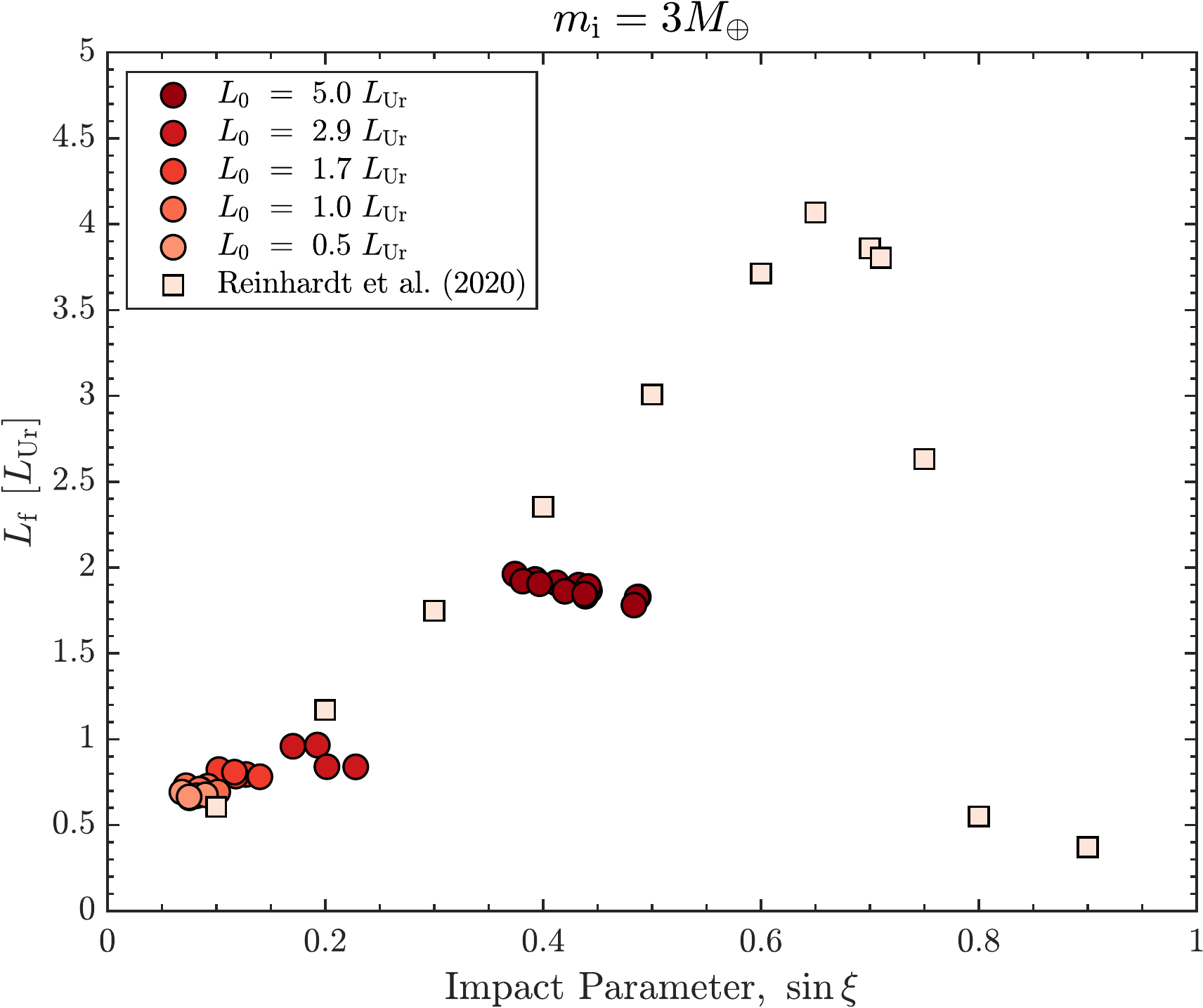}
    \caption{Post-impact AM of the planet+disk as a function of impact parameter for different initial planetary spins. As the pre-impact rotation is poorly constrained, the figure includes impacts between non-rotating planets and a differentiated $3M_{\oplus}$ impactor from \cite{reinhardt2020bifurcation} (squares), as well as results from simulations here in which the planet is rotating before the impact with a moderate obliquity (circles). We note that for high initial planetary rotations ($>2.9.0L_{\rm Ur}$), our impact simulations were setup such that the impact vector, $\vec{L}_{\rm i}$ is retrograde compared to the pre-impact planetary rotation, $\vec{L}_0$, hence the final AM in those simulations are smaller compared to the non-rotating cases from \cite{reinhardt2020bifurcation}.}
    \label{fig:CompAM}
\end{figure}

Previous works simulating giant impacts into Uranus \citep{Slattery1992,reinhardt2020bifurcation,kegerreis2018consequences, Woo20XX} have considered non-rotating target planets, and have used \textcolor{mod}{a minimum} post-impact planetary rotation as a proxy to define successful impact outcomes.  However, the post-impact planetary structure is significantly inflated, due to the high temperatures of the upper envelope and because the structure has not reached hydrostatic equilibrium within a few days after impact (e.g., Figure \ref{fig:Snapshot_Impact}). During planetary cooling and contraction, the rotation rate increases while the planet's rotational angular momentum (AM) remains constant.  Because of this, the planet's AM (which dominates the total system AM), rather than its rotation rate, is the needed proxy for determining whether a given post-impact system is consistent with the current Uranian system.  Previously defined ``successful'' impacts (e.g., with $\sin\xi \geq0.6$, $3M_\oplus$ impactor masses \textcolor{mod}{and $18-20\ {\rm km/s}$} impact velocities) produced disks with masses $\ge 10^{-2}M_{\rm Ur}$ \citep{reinhardt2020bifurcation,Woo20XX}. However,  such cases yield post-impact systems whose angular momenta are greatly in excess --by factors of 2 to 4 -- of that in the current Uranus system (see Figure \ref{fig:CompAM}). 
No means of extracting this AM after a Uranus-tipping impact has been demonstrated.  Rapid pre-impact target rotation in the opposite sense could be added to make $L_f \sim L_{\rm Ur}$.  However, our simulations (within circle in Fig. \ref{fig:DiskMass_Lf}a) imply that the disk mass would then be greatly reduced.  Thus the production of a massive disk with $\ge 10^{-2} M_{\rm Ur}$ appears inconsistent with the Uranus system AM. 

\begin{figure}
    \centering
   \includegraphics[width=0.45\linewidth]{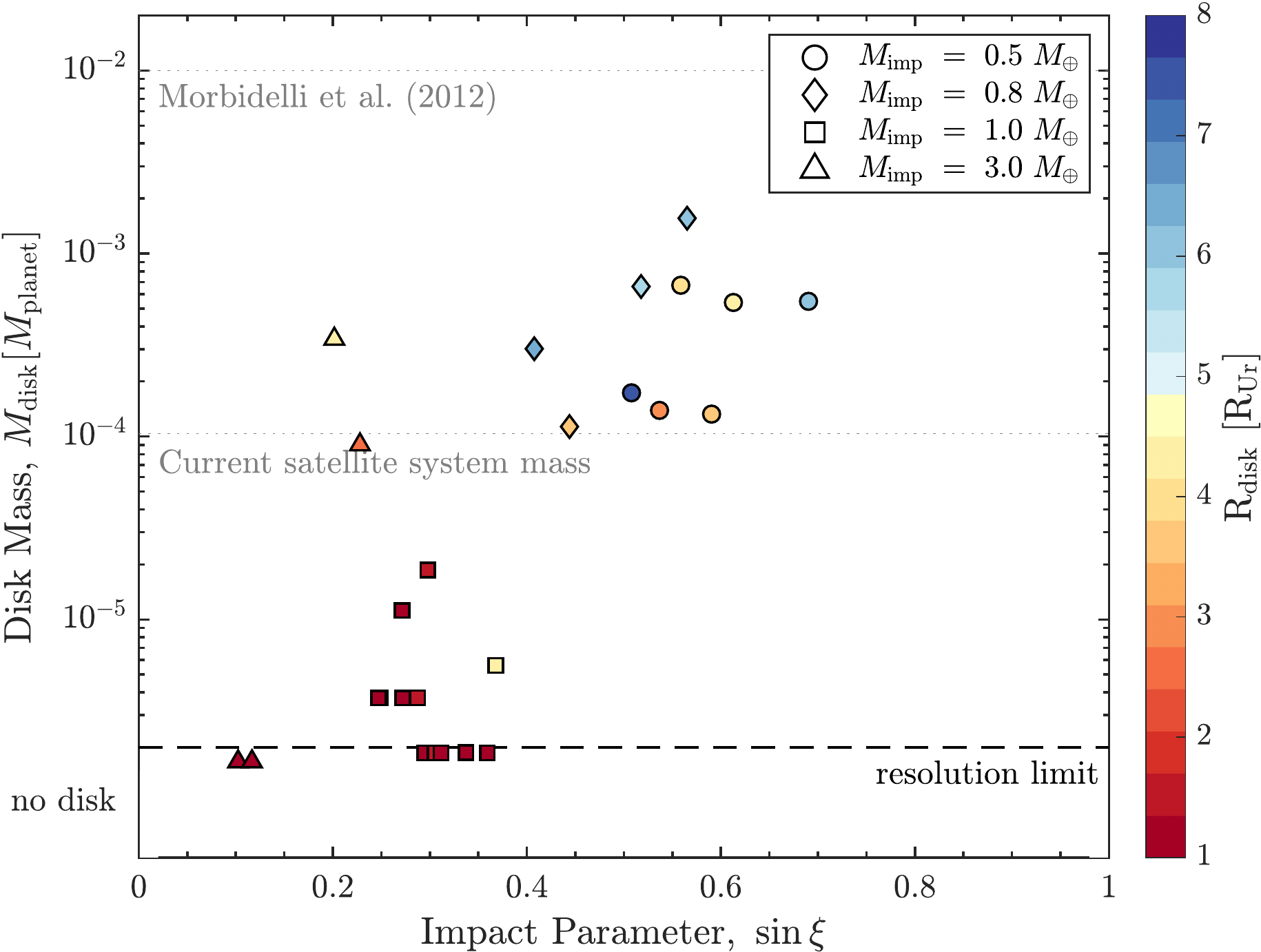}

    \caption{Disk mass as a function of the scaled impact parameter, $\sin{\xi}$, for $0.5$ (circle), $0.8$ (diamond), $1$ (square) and $3\, M_\oplus$ (triangles) impactor mass for simulations that resulted in a prograde disk and a final AM of $0.8-1.2 L_{\rm Ur}$. The colors of the markers correspond to the disk radius. Markers below the black dashed line in panel correspond to cases where no disk particles were detected. The upper grey dotted line represents the disk mass required to realign the outer disk to the planet's post-impact equatorial plane \citep{MORBIDELLI2012737}. The lower grey dotted line represents the mass of the current Uranian satellite system.}
    \label{fig:DiskMass_successful}
\end{figure}

In addition to its need for a massive impact-generated disk, the \cite{MORBIDELLI2012737} model also has specific requirements on the alignment of the giant impact relative to the pre-impact planet spin axis, and its assumption of a prior satellite system formed via gas co-accretion would seem to imply that the planet should have had a substantial spin before the impact as well.  In this work, we take these constraints into account, as well as the need for the Uranus-tipping impact to leave a system with an appropriate AM, when determining the allowed combinations of impactor mass, impact angle, pre-impact planet rotation, and the relative angle between the pre- and post-impact planet spin axes.  For example, for an impactor mass of $3 M_\oplus$, as shown in Figure \ref{fig:CompAM}, the allowed impact angles are restricted to almost head-on configurations (with $\sin\xi<0.26$), because larger impact angles yield systems with large AM excesses, \textcolor{mod}{ unless an extremely large pre-impact rotation in the opposite sense of the impact is included.}  We find that for configurations that meet the appropriate angular momentum criteria, the resulting impact-generated disk masses are $\sim 10^{-5}$ to $10^{-3}M_{\rm Ur}$ (Figure \ref{fig:DiskMass_successful}), much less than the $10^{-2}M_{\rm Ur}$ c-disk mass needed in the \cite{MORBIDELLI2012737} model.

\textit{We thus conclude that the \cite{MORBIDELLI2012737} model as originally proposed does not appear viable}.  However, there are substantial strengths of their overall co-accretion + giant impact hybrid view.  First, the only currently known explanation for Uranus' $98$ degree obliquity is a giant impact, and it seems that such an event must have occurred after the solar nebula dissipated.  If the giant impact instead occurred while the solar nebula was present, then even a small mass in nebular gas flowing into an accretion disk around Uranus would have destroyed the current moons, because the accretion disk gas would have had orbited the planet in the same sense as Uranus' orbit around the Sun, while the current moons orbit in the opposite direction.  This configuration would have led to the very rapid loss of the current moons due to inward gas drag \citep{Salmon20XXUranus}.  Gas co-accretion after a giant impact that occurred prior to nebular dissipation could have produced new satellites, but these would have orbited in the opposite sense to those observed today.  Thus there is strong circumstantial evidence that the Uranus-tipping giant impact occurred after the nebula had dissipated.  

Given this timing, it is probable that Uranus would have had a primordial satellite system at the time of the giant impact event, as it is thought that the accretion of satellites from a circumplanetary disk in the late stages of gas planet accretion would be common \citep{Canup_2006,ward2010circumplanetary,szulagyi2018situ}. \textcolor{mod}{ \cite{Woo20XX} argue that perhaps gas accretion onto Uranus did not yield a co-accretion disk and satellites because Uranus never accreted gas in a runaway mode.  Whether a disk forms during gas infall depends in part on how the radius of the planet compares to the so-called centrifugal radius of the infalling gas, i.e., the radius at which the gas would achieve keplerian orbit about the planet based on its specific angular momentum \citep{ward2010circumplanetary}. If the planet radius is larger than the centrifugal radius, gas is accreted directly onto the planet and there may be no circumplanetary disk \citep{ward2010circumplanetary}. A planet undergoing runaway gas accretion will be highly heated and distended, and a planet like Uranus that avoids runaway gas accretion may be more likely to be sufficiently cool and compact to have a co-accretion disk.} Thus the basic premise of the \cite{MORBIDELLI2012737} model -- that there would have been a system of prograde (with respect to Uranus' orbit) satellites formed by earlier gas co-accretion at the time of a Uranus-tipping giant impact -- remains compelling.  Indeed, it seems to these authors more probable than the alternative view that there were no prior satellites at the time of Uranus tipping giant impact, as assumed by models that seek to form the moons from a giant impact alone. Further, the difference in bulk composition between inner Miranda, which appears ice-rich, and the half rock, half ice compositions of all of the larger outer moons seems an important clue that they may have originated from different sources of material, with Miranda perhaps representing material associated with the Uranus-tipping giant impact, as suggested by \cite{Salmon20XXUranus}.

Accordingly, we suggest a variant on the co-accretion + giant impact concept. Consider an original prograde satellite system at Uranus that was more radially compact than the current Uranian satellites. \textcolor{mod}{ This configuration may be possible if the centrifugal radius for infalling gas was smaller during Uranus' gas accretion than during Jupiter and Saturn's final gas accretion, and/or if the location in the circumuranian disk where dust aggregates and forms satellite seeds was closer to the planet \citep{szulagyi2018situ}, compared to the location in the Jupiter's circumplanetary disk \citep{tanigawa2012distribution}.} Nodal randomization of the debris produced as this system collisionally destabilizes after a Uranus-tipping giant impact could then be achieved through forcing due to Uranus' $J_2$ in combination with a substantially less massive c-disk than considered in \cite{MORBIDELLI2012737}.  The Uranus-tipping giant impact might then produce only a low-mass, compact ice-rich inner disk, perhaps comparable to the few $\times 10^{-6}$ to $10^{-5}M_{\rm Ur}$ Roche-interior disk needed to later yield Miranda and the other inner small Uranian moons via viscous spreading and accretion near the Roche limit \citep{hesselbrock2017ongoing,Salmon20XXUranus}.  The challenges with the tendency for a massive inner \textit{c}-disk to destabilize and contaminate outer moons with ice-rich material identified in \cite{Salmon20XXUranus} would also be removed.

Debris from the compact prior satellite system would then be appropriately re-aligned with the planet's new equatorial plane.  However, how would this more compact disk of debris, orbiting within say $\sim 10R_{\rm Ur},$ yield a system of satellites out to Oberon at $23R_{\rm Ur}$?  We observe that circularization and realignment of the outer debris disk will generate prodigious heat.  Assuming the giant impact has tilted the planet by $\gtrsim50^\circ$, then the energy released during realignment of the debris disk is enough to vaporize all rock and ice in the primordial material \citep[eqn. 10 in][]{Nakajima2014259}; the energy is increased further if circularization is included.  Thus debris from the prior satellite system would be fully vaporized by its initial collisional evolution.  This fully vaporized material may then viscously expand prior to condensation and outer satellite re-accretion so as to yield the current satellites.  One model for how such viscous expansion might occur has been proposed by \cite{ida2020uranian}.  The accretion efficiency of the outer disk would need to be relatively high so as to preserve the $10^{-4}$ mass ratio that seems characteristic of systems formed by co-accretion.   Such a revised evolution is speculative, and determining whether it is viable will be a topic of our future work.


\appendix 
\section{Position and velocity of impactor}\label{Appendix:Li}

We set the post-impact AM to be comparable to the current system (magnitude $L_{\rm f}=L_{\rm Ur}$ and obliquity $\theta_{\rm f}=98^\circ$), and define the coordinate system such that $\vec{L}_{\rm f}$ lies in the $x$-$z$ plane (the azimuthal angle is $0^\circ$, see Figure \ref{fig:SchematicIS}).  Assuming that most of the impact AM is incorporated into the planet,  $\vec{L}_{\rm f}\approx\vec{L}_{0}+\vec{L}_{\rm i}$, gives:
\begin{equation}
\begin{aligned}
    \vec{L}_{\rm f} &= (L_{\rm f}\sin{\theta_{\rm f}},\ 0,\ L_{\rm f}\cos{\theta_{\rm f}})\\
    & = (L_0\sin{\theta_0}\cos{\phi_0}+L_{\rm i}\sin{\theta_{\rm i}}\cos{\phi_{\rm i}},\ L_0\sin{\theta_0}\sin{\phi_0}+L_{\rm i}\sin{\theta_{\rm i}}\sin{\phi_{\rm i}},\ L_{0}\cos{\theta_0}+L_{\rm i}\cos{\theta_{\rm i}})
\end{aligned}
\end{equation}
where $\theta$ [$\phi$] is the polar [azimuthal] angle and the i, $0$, f subscript indicate the impact AM, initial planetary AM and final AM, respectively. Given an initial planetary AM, $\vec{L}_{0}$, we can find the impact AM (magnitude and orientation) that would result in the current values of the system, $\vec{L}_{\rm f}$ using:

\begin{equation}
\begin{aligned}
    \tan{\phi_{\rm i}} &= \frac{L_0\sin{\theta_0}\sin{\phi_0}}{L_0\sin{\theta_0}\cos{\phi_0}-L_{\rm f}\sin{\theta_{\rm f}}}\\
    \tan{\theta_{\rm i}} &= \frac{L_0\sin{\theta_0}\sin{\phi_0}}{(L_0\cos{\theta_0}-L_{\rm f}\cos{\theta_{\rm f}})\cdot \sin{\phi_{\rm i}}}\\
    L_{\rm i} &= \frac{L_{\rm f}\cos{\theta_{\rm f}}-L_0\cos{\theta_0}}{\cos{\theta_{\rm i}}}
    \label{eq:L_i}
\end{aligned}
\end{equation}
Assuming that at the moment of impact the impactor (mass $m_{\rm i}$) is placed along the $y$-axis at $R_{\rm Ur}+r_{\rm i}$, the impactor velocity vector is:
\begin{equation}
    \vec{V}_{\rm i} = (V_{\rm{i},x},V_{\rm{i},y},V_{\rm{i},z})=\left(\frac{-L_{\rm i}\cos{\theta_{\rm i}}}{m_{\rm i}(R_{\rm Ur}+r_{\rm i})}, V_{\rm{i},y}, \frac{L_{\rm i}\sin{\theta_{\rm i}}\cos{\phi_{\rm i}}}{m_{\rm i}(R_{\rm Ur}+r_{\rm i})}\right)
\end{equation}
where $V_{\rm{i}, y}$ is defined by the chosen impact velocity magnitude, $V_{\rm i}$, such that $V_{\rm{i}, y}=\sqrt{V_{\rm i}-V_{\rm{i}, x}^2-V_{\rm{i}, z}^2}$.

\section{Stability of a vapor rich disk}\label{Appendix:VaporDisk}
The resulting disks are vapor-rich (vapor fraction $>90\%$), hence the assumption that disk particles follow Keplerian orbits is an oversimplification. We will show in this section that the resulting pressure supported disks may become unstable, decreasing the disk mass even further than estimated in the main text.

The specific disk AM in the $z$-direction, $l_z$, of a pressure supported disk is \cite[e.g.,][]{Nakajima2014259}:
\begin{equation}
   l_z^2(r_{\rm xy}) = GM_{\rm planet}\cdot r_{\rm xy}+\frac{r_{\rm xy}^3}{\rho_*(r_{\rm xy})}\frac{dP}{dr_{\rm xy}}
   \label{eq:lz}
\end{equation}
where $r_{\rm xy}^2\equiv x^2+y^2$ is the distance from the rotational axis, $\rho_*$ is the mid-plane density and $P$ is the pressure.
We assume that the disk is isothermal ($T$=const) and in hydrostatic equilibrium in the vertical direction such that the density at level $z$ above the mid plane is \citep{lyra2019initial}:
\begin{equation}
    \rho(r_{\rm xy},z)=\rho_*(r_{\rm xy})e^{-z^2/2H^2}
\end{equation}
where $H=c_s/\Omega$ is the scale height, $c_s$ the sound speed and $\Omega$ the orbital frequency. We simplify the mid plane density distribution using a power, law $\rho_*\propto r_{\rm xy}^{-q_\rho}$ \citep{lyra2019initial}, where $q_\rho$ is obtained by conserving the disk mass and AM (i.e., $\int_{R_{\rm planet}}^{R_{\rm disk}}\sigma(r_{\rm xy}) r_{\rm xy} dr_{\rm xy} =M_{\rm disk}$ and $\int_{R_{\rm planet}}^{R_{\rm disk}}\sigma(r_{\rm xy}) r_{\rm xy}^2 \Omega(r_{\rm xy}) dr_{\rm xy}=l_z M_{\rm disk}$, where $R_{\rm disk}$ is the extent of the disk and $\sigma\sim\rho/H$ is the surface density, \citealp[e.g.,][]{ward2011vertical}).
Using the ideal gas law, $P=\rho c_s^2$, the pressure gradient is:
\begin{equation}
    \frac{1}{\rho_*}\frac{dP}{dr_{\rm xy}}=-\frac{1}{r_{\rm xy}}q_{\rho}c_s^2
    \label{eq:dPdr}
\end{equation}
where $c_s=\sqrt{R T/\mu}$, $R$ is the gas constant, and $\mu$ is the mean molecular weight (calculated using the resulting water-to-hydrogen fraction in each disk).

In the proto-lunar disk the energy released during circularization of the impact-generated debris disk, $\Delta E$, is insignificant compared to the post-impact thermal energy \citep{Nakajima2014259}. However in a circum-Uranian disk, due to the more massive center body, the circularization energy is comparable to or larger than the post-impact thermal energy. Therefore, to estimate the temperature (and, hence, the sound speed), we assume that $\bar{T}\sim \Delta E/c_v$ (where $c_v$ is the specific heat, $2\times10^8\ \rm{[erg/g/K]}$ for water and $1.3\times10^7\ \rm{[erg/g/K]}$ for hydrogen). 

\begin{figure}
     \centering
    \includegraphics[width=0.45\linewidth]{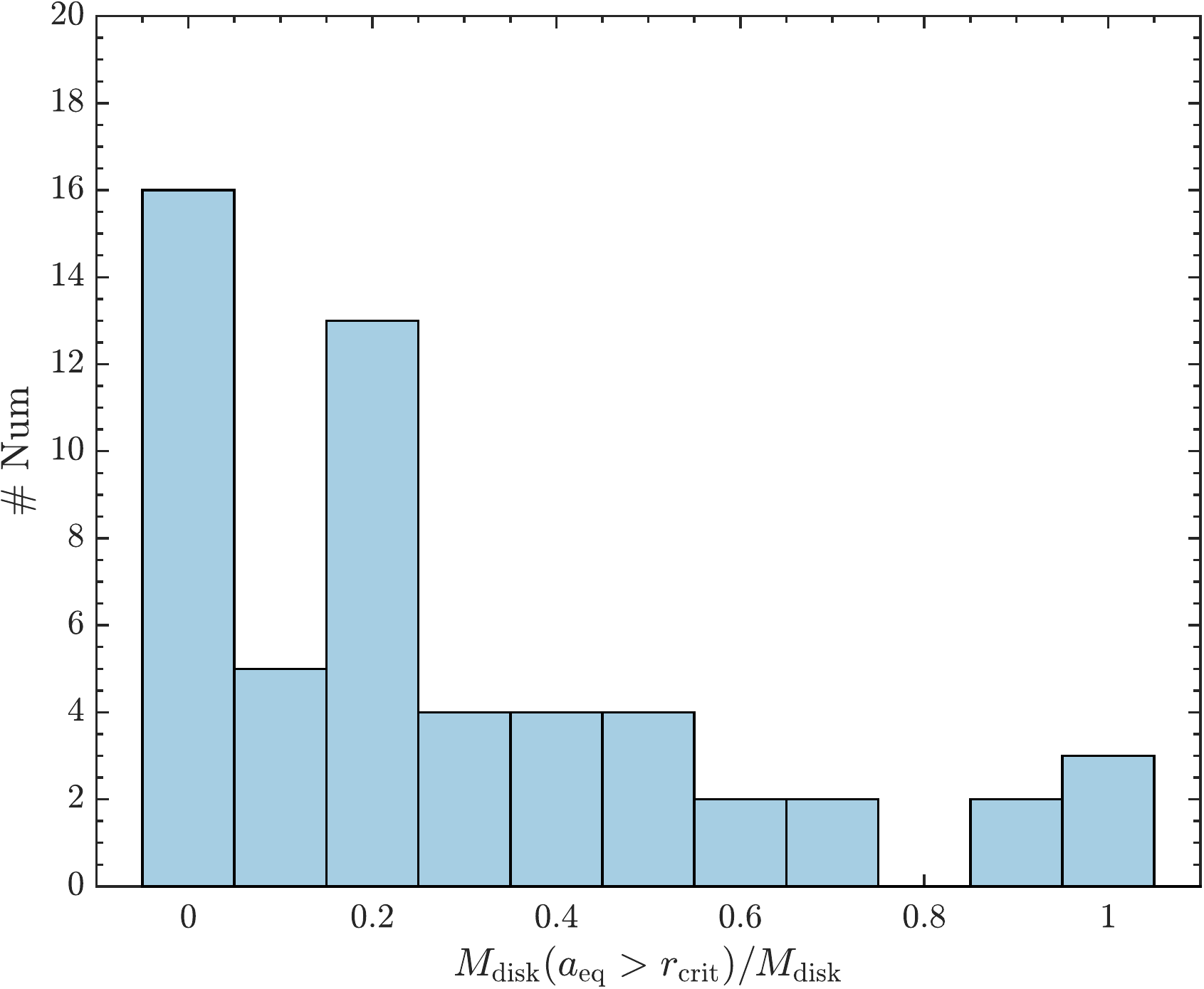}
     \caption{Distribution of the disk mass fraction that is beyond the critical radius, $r_{\rm disk}$. $50\%$ of the resulting disks have a substantial fraction of their mass ($>0.2$) that is not dynamically stable ($dl_z/dr_{\rm{ xy}}<0$), hence radial redistribution will occur. In this figure we exclude disks with $<2\times 10^{-5}M_{\rm Ur}$ ($<10$ particles).}
     \label{fig:HistRadiusCrit}
\end{figure}

The Rayleigh stability criterion requires that the AM of a disk will increase outward $dl_z/dr_{\rm xy}>0$ \citep{chandrasekhar2013hydrodynamic}. E.g., for a Keplerian flow, $l_z=\Omega_k r_{\rm xy}^2=\sqrt{GM_{\rm planet} r_{\rm xy}}$, hence an undisturbed Keplerian flow is unconditionally stable. Using eqn. (\ref{eq:lz}) and (\ref{eq:dPdr}) we can show that for the pressure supported disk defined above, at a critical distance,  $r_{\rm crit}$, the Rayleigh stability criterion is no longer valid:
\begin{equation}
    r_{\rm crit}=\frac{1}{2}\frac{GM_{\rm planet}}{q_{\rho}c_s^2}
\end{equation}
Beyond this distance radial redistribution will occur while a significant portion of the vapor could migrate inward onto the planet. Some of the resulting $r_{\rm crit}(q_\rho,c_s)$ are smaller than the radial extend of the debris disks, placing significant parts of their mass beyond the Rayleigh stability limit (Figure \ref{fig:HistRadiusCrit}). Hence, the orbiting disk mass estimated in section \ref{sec:DiskMass} is an upper limit to the expected orbiting disk mass.


\newpage
\acknowledgments
This research was supported by NASA’s Emerging Worlds (80NSSC18K0733) program.  RR is an Awardee of  the  Weizmann  Institute  of  Science - National  Postdoctoral  Award  Program  for Advancing  Women  in  Science. We thank Christian Reinhardt  for providing the angular momentum values from their simulations (Reinhardt et al. 2020) and for the helpful discussion. \textcolor{mod}{We thank Nadine Nettelmann and an anonymous reviewer for their thoughtful comments and suggestions that improved the final version of this manuscript.}
\bibliographystyle{aasjournal}

\bibliography{AllBib}{}



\end{document}